\documentclass{elsart}
\journal{} 

\usepackage{epsfig}
\usepackage{amssymb}

\begin{document}
\begin{frontmatter}

\title{Identification of photons in double beta-decay experiments 
using segmented germanium detectors - studies with a GERDA Phase~II
prototype detector}
\author{I.~Abt}, 
\author{A.~Caldwell}, 
\author{K.~Kr\"oninger\corauthref{cor}}\ead{kroening@mppmu.mpg.de}, 
\author{J.~Liu}, 
\author{X.~Liu}, 
\author{B.~Majorovits} 
\address{Max-Planck-Institut f\"ur Physik, M\"unchen, Germany}
\corauth[cor]{Max-Planck-Institut f\"ur Physik, M\"unchen, Germany, 
Tel. +49-(0)89-32354-337}

\begin{abstract} 
The sensitivity of experiments searching for neutrinoless double
beta-decay of germanium was so far limited by the background induced
by external $\gamma$-radiation. Segmented germanium detectors can be
used to identify photons and thus reduce this background component. \\
The GERmanium Detector Array, {\sc GERDA}, will use highly segmented
germanium detectors in its second phase. The identification of
photonic events is investigated using a prototype detector. The
results are compared with Monte Carlo data. \\
\end{abstract} 
\end{frontmatter} 


\section{Introduction}
\label{section:introduction}

Neutrinoless double beta-decay ($0\nu\beta\beta$) is expected to
occur, if the neutrino is a massive Majorana particle. The observation
of the $0\nu\beta\beta$-process would not only reveal the nature of
the neutrino as a Majorana particle but could also provide information
about the absolute neutrino mass scale (see, e.g.~\cite{zeronubb}). \\

The germanium isotope $^{76}$Ge is a prominent candidate for the
observation of the $0\nu\beta\beta$-process. Experiments searching for
neutrinoless double beta-decay of $^{76}$Ge use high purity germanium
detectors as source and detector simultaneously. Their sensitivity is
limited by unidentified background events which in previous
experiments were mostly induced by external $\gamma$-radiation.  The
Heidelberg-Moscow and IGEX experiments set 90\%~C.L. lower limits on
the half-life of the process of
$T_{1/2}>1.9\cdot10^{25}$~years~\cite{HM} and
$T_{1/2}>1.6\cdot10^{25}$~years~\cite{IGEX}, respectively. An evidence
for the observation of the $0\nu\beta\beta$-process was claimed by
parts of the Heidelberg-Moscow collaboration with
$T_{1/2}=1.2\cdot10^{25}$~years~\cite{Klapdor}. \\
 
The GERmanium Detector Array, {\sc GERDA}~\cite{proposal}, is a new
germanium double beta-decay experiment being installed in Hall A of
the INFN Gran Sasso National Laboratory (LNGS), Italy. Its main design
feature is to operate germanium detectors directly in liquid argon
which serves as cooling medium and as a shield against external
$\gamma$-radiation. A detailed description of the experiment can be
found in~\cite{proposal,segmentation}. \\

The detectors for the second phase of the experiment (Phase~II) will
be enriched in $^{76}$Ge to a level of about 86\% and will have a mass
of approximately 2~kg each. For the first time, highly segmented
germanium detectors will be used in a double beta-decay
experiment. The segmentation scheme is chosen to minimize the
background level in the energy region around
$Q_{\beta\beta}=2\,039$~keV. The current detector design foresees a
6-fold segmentation in the azimuthal angle $\phi$ and a 3-fold
segmentation in the height $z$. All segments and the core are read out
separately to allow a better identification of photons. The estimated
gain in background reduction for the {\sc GERDA} experiment is
discussed in~\cite{segmentation}. \\
 
In this paper the results of a study with a {\sc GERDA} Phase~II
prototype detector are presented. The identification of events with
photons in the final state is
investigated. Section~\ref{section:background} summarizes the photon
identification using coincidences between segments. The underlying
physics processes and their signatures are described as are the event
selection and the analysis strategy. Section~\ref{section:setup}
describes the experimental setup of the prototype detector and the
data sets collected. The Monte Carlo simulation is introduced in
section~\ref{section:simulation}. The results of the study and
comparisons with Monte Carlo data are given in
section~\ref{section:results}. Section~\ref{section:conclusions}
concludes and discusses the significance for the {\sc GERDA}
experiment.


\section{Identification of photon events using segment coincidences}
\label{section:background}

The volume over which energy in a single event is deposited inside a
detector depends on the incident particles. Segmented detectors can be
used to identify events with photons in the final state by requiring
coincidences between the segments of a detector. This technique is
well established in nuclear experiments such as {\sc
AGATA}~\cite{agata} and {\sc GRETA}~\cite{greta}, and provides a basis
for $\gamma$-ray tracking~\cite{gammaraytracking}. \\
 
The potential of segmented detectors for double beta-decay experiments
has also been investigated by the {\sc Majorana}
collaboration~\cite{majorana} using a clover detector, consisting of
four detectors with two longitudinal segments each, and Monte Carlo
simulations~\cite{segmentationmc}. \\

\subsection{Signatures and physics processes} 
 
The signatures of events encountered in double beta-decay experiments
can be classified according to the particles in the final state. A
detailed classification for these events is given
in~\cite{segmentation}. For the identification of photon events only
two such classes are considered here:

\begin{itemize} 
\item Class~L: Local energy deposit. Three different types of events are 
part of this class: (a) Events with only electrons in the final state.
Electrons in the MeV-energy region have a range of the order of a
millimeter in germanium~\cite{segmentation,firestone}. Energy is
therefore deposited locally. Double beta-decay events which have two
electrons in the final state are of this type (these events correspond
to Class~I events in reference~~\cite{segmentation}).  (b) Events with
photons in the final state in which a photon Compton-scatters only
once inside the fiducial volume of the detector. Energy is thus
deposited locally. (c) 'Double escape' events: If a photon produces an
electron-positron pair and both photons from the subsequent
annihilation escape, energy is deposited locally.
\item Class~M: Multiple energy deposits. Photons emitted in radioactive 
decays have energies in the MeV-energy region and interact dominantly
through Compton scattering in germanium. The range of these photons is
of the order of centimeters. The different interactions are separated
by distances large compared to the scale of Class~L events. This class
is composed of Classes~II-IV in reference~\cite{segmentation}.
\end{itemize} 

It should be noted that with the technique presented in this paper the
three event types in Class~L cannot be separated but only be
distinguished from Class~M events.

\subsection{Event selection and identification of photon events} 
  
Due to the well separated multiple energy deposits Class~M events are
expected to deposit energy predominantly in more than one
segment. Class~L events will predominantly deposit energy in only one
segment. Events in which more than one segment measures deposited
energy can thus be identified as Class~M events. 


\section{Experimental setup and data sets} 
\label{section:setup}

\subsection{Experimental setup} 

The {\sc GERDA} Phase~II prototype detector under study is a high
purity $n$-type germanium crystal with a true coaxial geometry. It is
70~mm high and has an outer diameter of 75~mm. The inner diameter is
10~mm. The detector is 6-fold segmented in the azimuthal angle $\phi$
and 3-fold segmented in the height $z$. It is placed inside a
two-walled aluminum cryostat with a combined thickness of 6~mm. The
operation voltage of the detector is (+)$3\,000$~V. \\

\begin{figure}[ht!] 
\center 
\epsfig{file=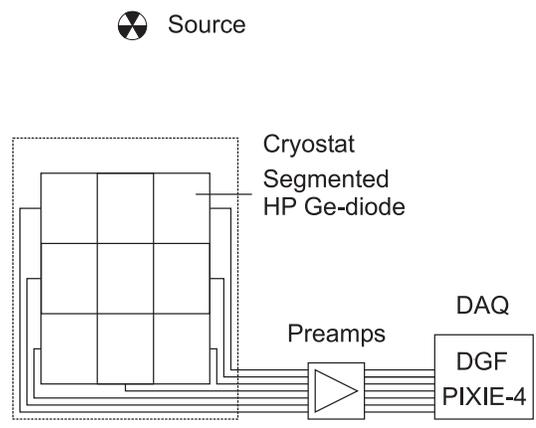,width=0.5\textwidth}
\caption{Schematic diagram of the detector and the experimental setup. 
\label{fig:setup}}
\end{figure}

A schematic diagram of the detector and the experimental setup is
given in Figure~\ref{fig:setup}. The core and each segment are read
out using charge sensitive PSC~823 pre-amplifiers. The pre-amplified
signals are digitized using a data acquisition system based on 5
14-bit ADC PIXIE-4 modules at a sampling rate of 75~MHz. In this
configuration the energy resolution of the core is approximately
2.6~keV (FWHM at $1\,333$~keV), the energy resolution of the segments
varies between 2.4~keV and 4.7~keV with an average segment energy
resolution of 3.3~keV. The threshold of the core and the segments was
set to 20~keV. Cross-talk between the core and the segment
pre-amplifiers and a constraint in the DAQ system, which resulted in
the inability to handle late arriving signals, caused a fraction of
less than 10\% of individual segment signals to not be recorded. \\
 
A detailed description of the setup and the prototype properties will
be published~\cite{characterization}.

\clearpage 

\subsection{Measurements and data sets} 
 
Several measurements were performed with different radioactive sources
positioned 10~cm above the center of the detector. Energy and time
information in all segments and the core were recorded on an
event-by-event basis. An event was recorded, if the energy measured in
the core exceeded the threshold. Measurements were performed with
three different sources: (1) a 60~kBq $^{60}$Co source, (2) a 100~kBq
$^{228}$Th source and (3) a 75~kBq $^{152}$Eu source. The
corresponding data sets are referred to as ``source data sets'' in the
following and contain approximately $4\cdot10^{6}$~events each. An
additional measurement without any source was performed in order to
estimate the background in the laboratory. This ``background data
set'' contains approximately $10^{6}$ events. \\

\begin{figure}[ht!]
\center
\begin{tabular}{cc}
\begin{minipage}[ht!]{0.40\textwidth}
\mbox{\epsfig{file=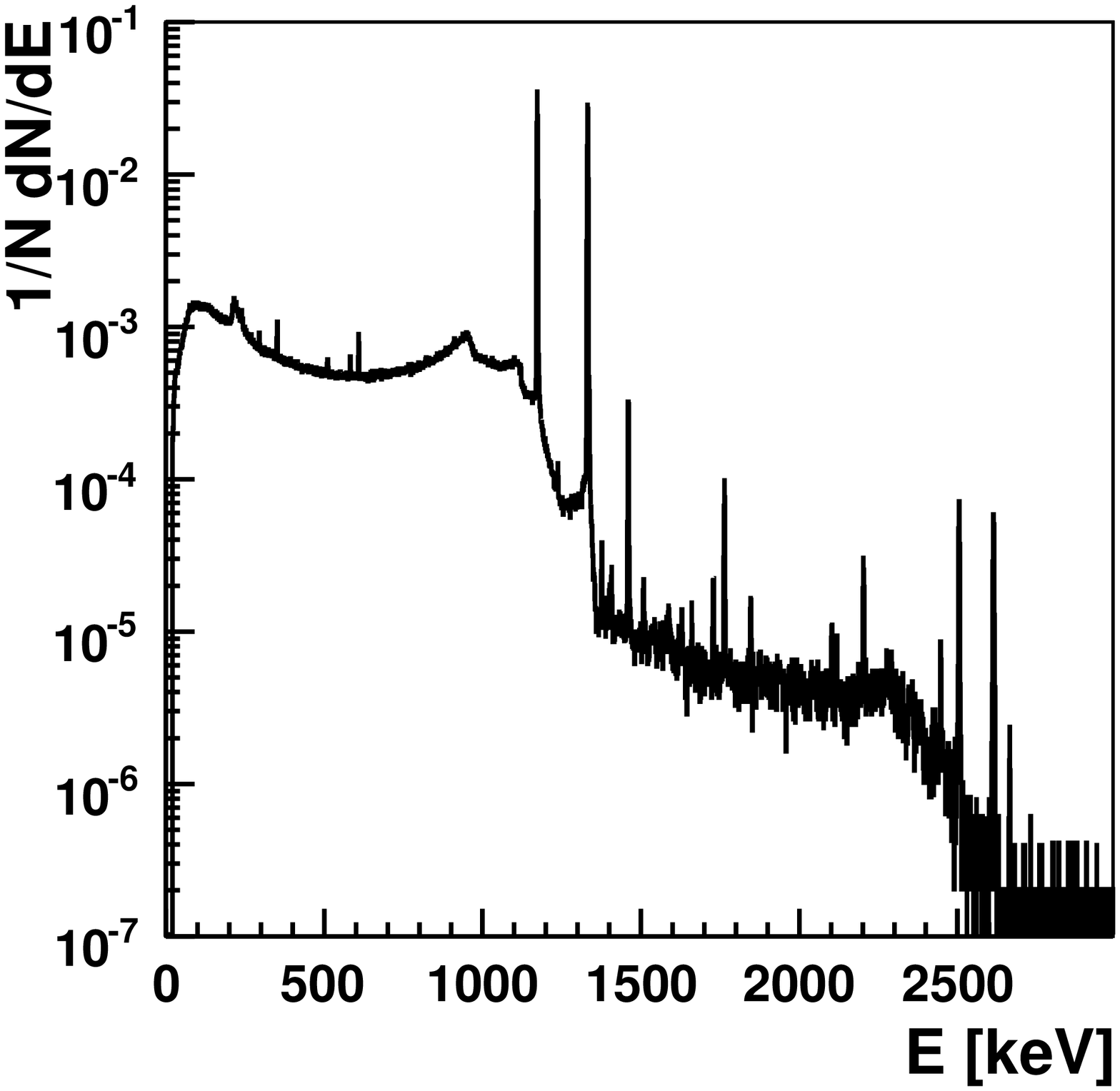,width=\textwidth}}
\end{minipage}
&
\begin{minipage}[ht!]{0.40\textwidth}
\mbox{\epsfig{file=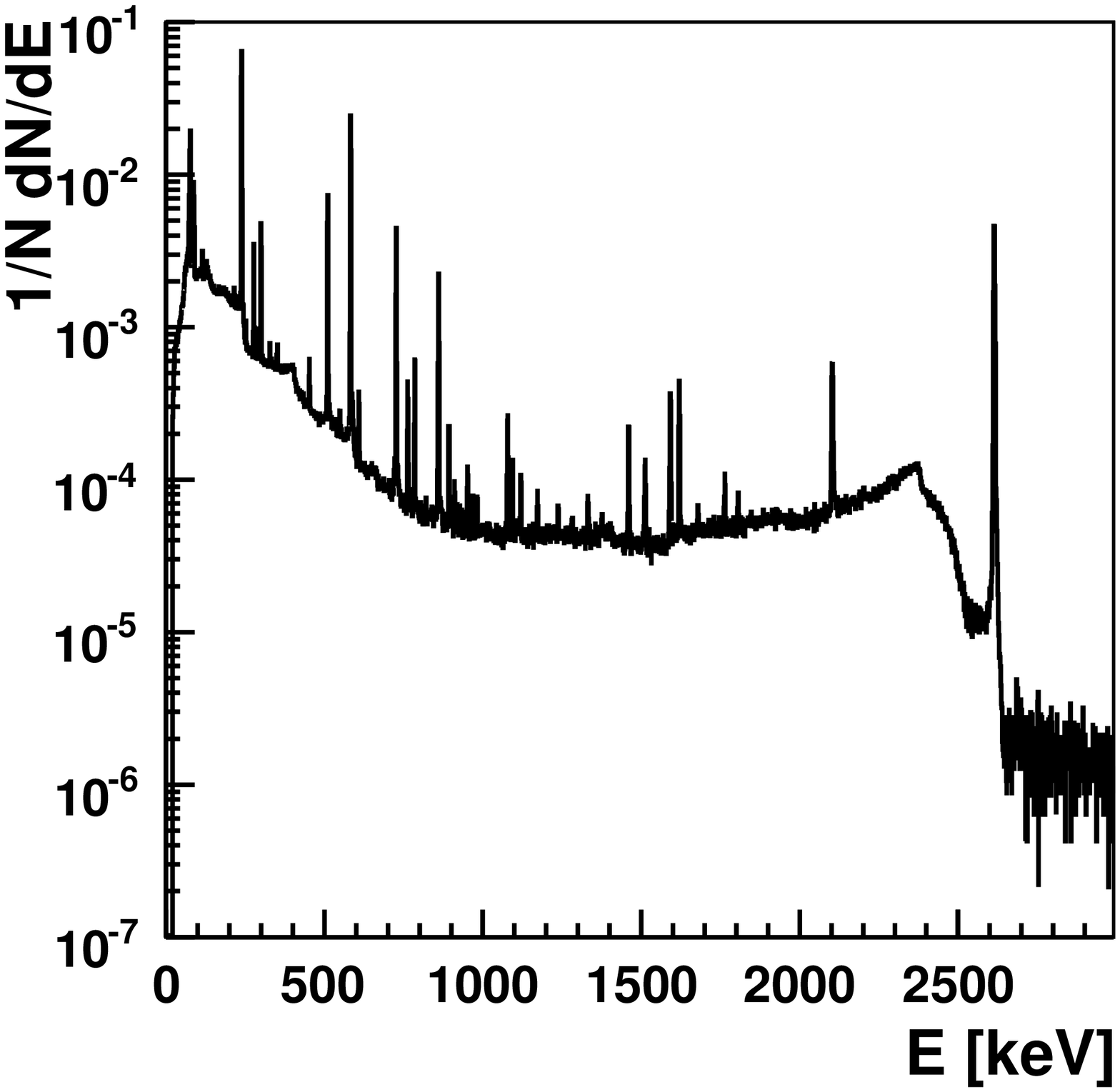,width=\textwidth}}
\end{minipage} \\
\\
\begin{minipage}[ht!]{0.40\textwidth}
\mbox{\epsfig{file=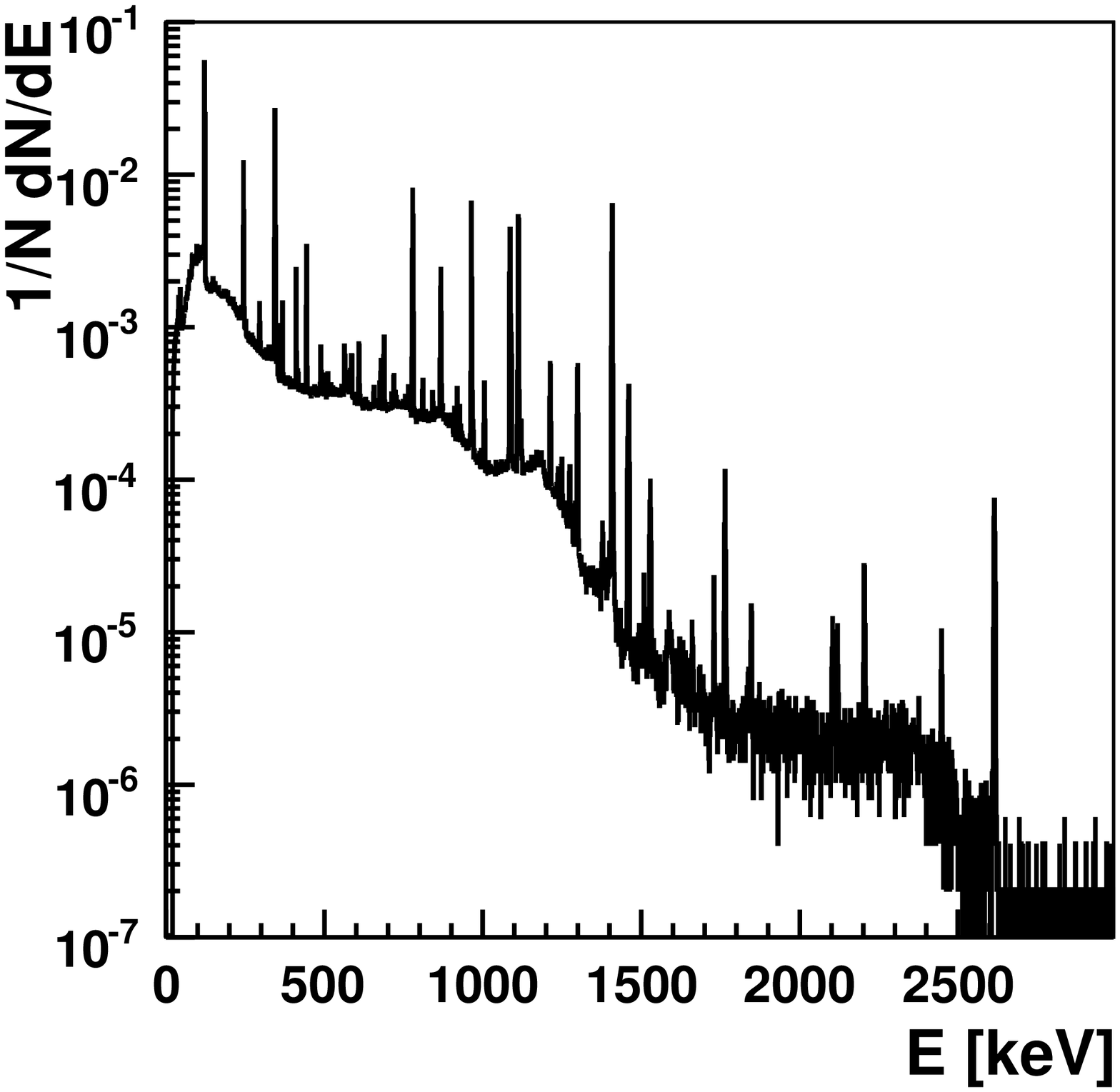,width=\textwidth}}
\end{minipage}
&
\begin{minipage}[ht!]{0.40\textwidth}
\mbox{\epsfig{file=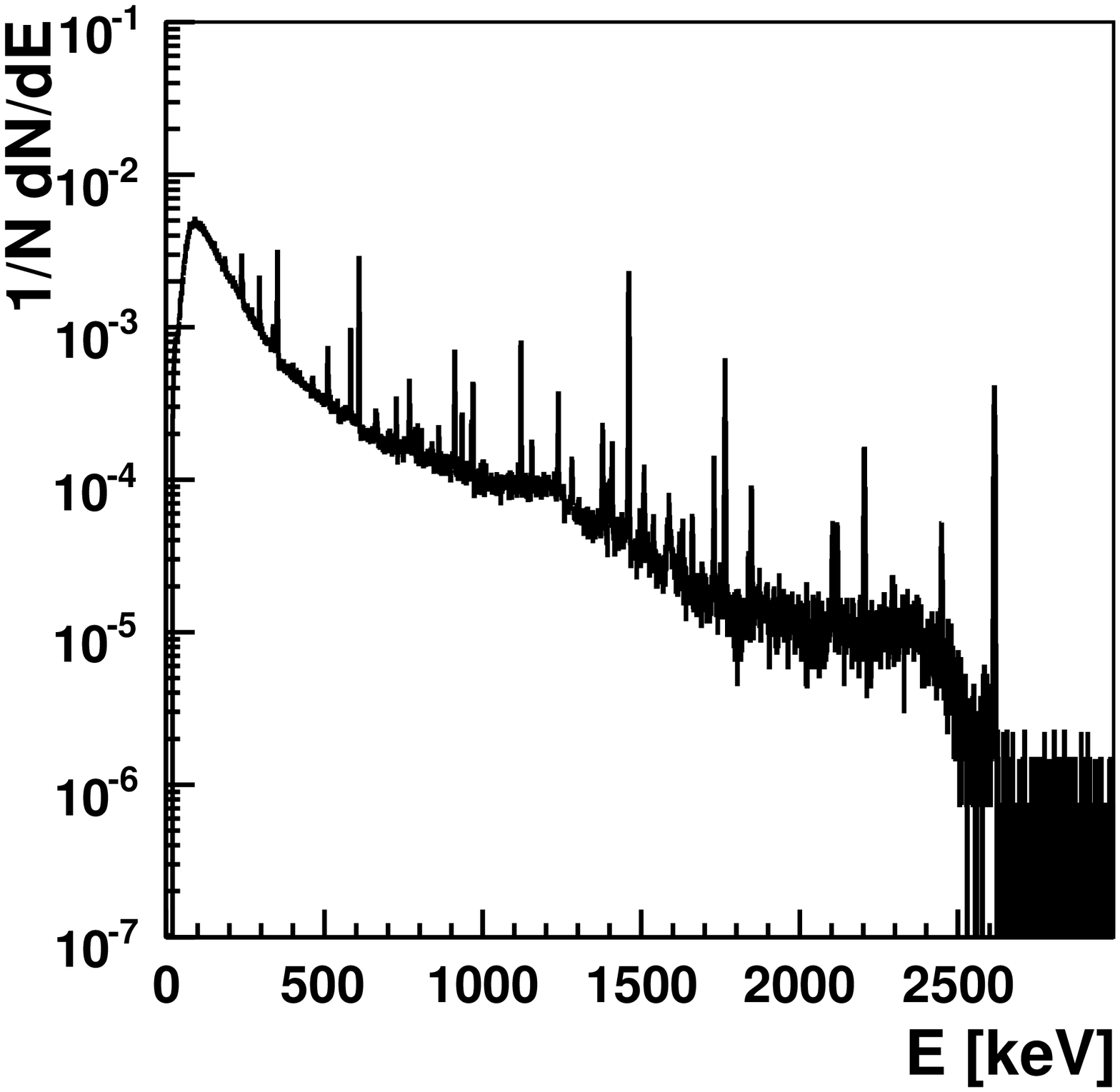,width=\textwidth}}
\end{minipage} \\
\end{tabular}
\caption{Raw energy spectra obtained with the core electrode for the 
$^{60}$Co (top, left), $^{228}$Th (top, right) and the $^{152}$Eu
(bottom, left) source data sets. The sources were placed 10~cm above
the detector. The energy spectrum for the background data set is also
shown (bottom, right). The binning is 1~keV and the spectra are
normalized to the area. 
\label{fig:rawspectra}}
\end{figure}

Figure~\ref{fig:rawspectra} shows the raw energy spectra obtained with
the core electrode for the three source data sets and the background
data set.

\clearpage 


\section{Monte Carlo simulation} 
\label{section:simulation} 

A Monte Carlo simulation of the prototype setup was performed using
the GEANT4~\cite{Agostinelli:2002hh} based {\sc MaGe}
framework~\cite{MaGe}. The energy deposited in each segment is
recorded and the core energy is calculated by adding all segment
energies. \\

The drift anisotropy of charge carriers inside the germanium
diode~\cite{RR} can cause electrons and holes to deviate from their
drift path. It is therefore possible to measure energy in one segment
even if the energy was deposited in the neighboring segment. Hence, a
correction is applied to the segment energies. An effective model is
used which assigns a segment to each energy deposit depending on its
position with respect to the axes of the crystal and the segment
borders. The maximum angular shift is 3.5$^{\circ}$. The directions of
the crystal axes were measured and used as input for the Monte
Carlo. This includes an overall variation of this effect by 40\% with
respect to the two hemispheres. \\

Each segment is assigned a relative efficiency with respect to the
core on the order of 90\%. This effectively models the
DAQ-inefficiency. The segment and core energies are individually
smeared according to the energy resolution of the prototype detector
measured in each channel. \\


\section{Results} 
\label{section:results} 

The results of the measurements are presented in the following and
compared to Monte Carlo data. In order to account for background from
radioactive isotopes in the laboratory the fraction of background
events in each data set is estimated.
 
\subsection{Background estimate} 

The number of background events in a given source data set is
estimated using characteristic photon lines in the spectrum. These
lines are associated with the decays of $^{214}$Pb (352~keV),
$^{214}$Bi (609~keV, $1\,120$~keV, $1\,765$~keV, $2\,204$~keV) and
$^{40}$K ($1\,461$~keV). The photon lines are fitted with a Gaussian
plus linear function and the number of events, $n_{i}$, under each
peak is calculated. \\
 
For the background data set, the fraction of events under the $i$th
peak is denoted $f_{i} = N_{i}/N$, where $N_{i}$ is the number of
events under the $i$th peak and $N$ is the total number of events in
the spectrum. For each source data set the total number of background
events, $n_{\mathrm{bkg}}$, is estimated by minimizing a
$\chi^{2}$-function defined as

\begin{equation} 
\chi^{2} = \chi^{2}(n_{\mathrm{bkg}}) = \sum_{i} \frac{\left(n_{\mathrm{bkg}} \cdot f_{i} - n_{i} \right)^{2}}{\sigma_{i}^{2} + n_{i}}, 
\end{equation} 

\noindent 
where $\sigma_{i}$ is the Poissonian uncertainty on the
expression $n_{\mathrm{bkg}} \cdot f_{i}$. \\

The fraction of background events in the source data sets are
estimated as 14.0\% ($\chi^{2}/d.o.f.=1.0$) for $^{60}$Co, 8.3\%
($\chi^{2}/d.o.f.=0.7$) for $^{228}$Th and 15.8\%
($\chi^{2}/d.o.f.=8.4$) for $^{152}$Eu. An uncertainty on the
background fraction of 0.1\% is estimated.

\subsection{Photon identification and reduction} 
\label{subsection:photonidentification} 

The segment multiplicity, $N_{\mathrm{S}}$, is defined as the number
of segments with measured energies larger than the threshold of
20~keV. \\

A measure for the identification of photonic events is the
\emph{number} suppression factor, $SF_{N}$, defined as the
ratio of the number of events within a 10~keV region around a certain
energy and the number of events which, in addition, have a segment
multiplicity of $N_{\mathrm{S}}=1$. In order to quantify the
identification of photons which deposit their full energy within the
detector the \emph{line} suppression factor, $SF_{L}$, is defined
similarly to $SF_{N}$, but the number of events is replaced by the
number of events under the photon peak under study. \\

For the calculation of the suppression factors the number of events in
the source data sets are corrected for the background by subtracting
the background contribution. \\

The number suppression factor is calculated for the $Q_{\beta\beta}$
region of $^{76}$Ge ($2\,039$~keV) whereas the line suppression
factors are calculated for the photon lines of $^{60}$Co
($1\,173$~keV, $1\,333$~keV and the summation peak at $2\,506$~keV),
$^{208}$Tl (511~keV, 583~keV, 861~keV, $2\,615$~keV and the
corresponding single and double escape peaks at $2\,104$~keV and
$1\,593$~keV), $^{212}$Bi ($1\,620$~keV) and $^{152}$Eu (122~keV,
245~keV, 344~keV, 779~keV, 964~keV, $1\,086$~keV, $1\,112$~keV and
$1\,408$~keV). The results are displayed in
Table~\ref{table:suppression} for data and Monte Carlo data. For a
discussion of the agreement between data and Monte Carlo see
section~\ref{subsection:datatomc}.

\begin{table}[ht!]
\caption{Suppression factors for different sources and energies
calculated for data and Monte Carlo. The background has been
subtracted from the data. For a discussion on the agreement between
data and Monte Carlo see section~\ref{subsection:datatomc}. The
uncertainties are statistical uncertainties only.
\label{table:suppression}}
\center
\begin{tabular}{lccccc}
\hline 
Source     & Energy             & $SF_{N}$ (data)                          & $SF_{L}$ (data)                        & $SF_{N}$ (MC)   & $SF_{L}$ (MC) \\
           & [keV]              &                                          &                                        &                 & \\ 
\hline												     
$^{60}$Co  & $1\,173$           &  -                                       & \phantom{0}2.56 $\pm$ 0.01\phantom{0}  & -               & \phantom{0}2.56 $\pm$ \phantom{0}0.01\phantom{0} \\ 
           & $1\,333$           &  -                                       & \phantom{0}2.63 $\pm$ 0.01\phantom{0}  & -               & \phantom{0}2.63 $\pm$ \phantom{0}0.01\phantom{0} \\ 
           & $2\,506$           &  -                                       & 34.6\phantom{0} $\pm$ 5.7\phantom{00}  & -               & 43.0\phantom{0} $\pm$ 11.0\phantom{00} \\ 
\hline 												     
           & $2\,039$           & 14.2\phantom{0} $\pm$ 2.1\phantom{0}     & -                                      & 12.5 $\pm$ 2.1  & - \\ 
\hline 												     
$^{228}$Th & \phantom{0\,}$511$ & -                                        & \phantom{0}1.92 $\pm$ 0.01\phantom{0}  & -               & 1.91 $\pm$ \phantom{0}0.02\phantom{0} \\ 
           & \phantom{0\,}$583$ & -                                        & \phantom{0}2.04 $\pm$ 0.01\phantom{0}  & -               & 2.01 $\pm$ \phantom{0}0.01\phantom{0} \\ 
	   & \phantom{0\,}$861$ & -                                        & \phantom{0}2.35 $\pm$ 0.03\phantom{0}  & -               & 2.37 $\pm$ \phantom{0}0.05\phantom{0} \\ 
           & $1\,593$           & -                                        & \phantom{0}1.09 $\pm$ 0.02\phantom{0}  & -               & 1.09 $\pm$ \phantom{0}0.04\phantom{0}\\ 
           & $1\,620$           & -                                        & \phantom{0}2.85 $\pm$ 0.01\phantom{0}  & -               & 2.84 $\pm$ \phantom{0}0.13\phantom{0} \\ 
	   & $2\,104$           & -                                        & \phantom{0}3.13 $\pm$ 0.01\phantom{0}  & -               & 3.20 $\pm$ \phantom{0}0.11\phantom{0} \\ 
           & $2\,615$           & -                                        & \phantom{0}3.04 $\pm$ 0.02\phantom{0}  & -               & 3.23 $\pm$ \phantom{0}0.04\phantom{0} \\ 
\hline 												     
           & $2\,039$           & \phantom{0}1.68 $\pm$ 0.02               & -                                      & 1.66 $\pm$ 0.05 & - \\ 
\hline												     
$^{152}$Eu & \phantom{0\,}$122$ & -                                        & \phantom{0}1.01 $\pm$ 0.002            & -               & 1.01 $\pm$ \phantom{0}0.003 \\ 
           & \phantom{0\,}$245$ & -                                        & \phantom{0}1.26 $\pm$ 0.01\phantom{0}  & -               & 1.22 $\pm$ \phantom{0}0.01\phantom{0} \\ 
           & \phantom{0\,}$344$ & -                                        & \phantom{0}1.54 $\pm$ 0.01\phantom{0}  & -               & 1.55 $\pm$ \phantom{0}0.01\phantom{0} \\ 
           & \phantom{0\,}$779$ & -                                        & \phantom{0}2.29 $\pm$ 0.01\phantom{0}  & -               & 2.26 $\pm$ \phantom{0}0.02\phantom{0} \\ 
           & \phantom{0\,}$964$ & -                                        & \phantom{0}2.46 $\pm$ 0.02\phantom{0}  & -               & 2.41 $\pm$ \phantom{0}0.02\phantom{0} \\ 
	   & $1\,086$           & -                                        & \phantom{0}2.54 $\pm$ 0.02\phantom{0}  & -               & 2.50 $\pm$ \phantom{0}0.03\phantom{0} \\ 
           & $1\,112$           & -                                        & \phantom{0}2.52 $\pm$ 0.02\phantom{0}  & -               & 2.54 $\pm$ \phantom{0}0.04\phantom{0} \\ 
           & $1\,408$           & -                                        & \phantom{0}2.64 $\pm$ 0.02\phantom{0}  & -               & 2.72 $\pm$ \phantom{0}0.02\phantom{0} \\ 
\hline
\end{tabular} 
\end{table} 

\begin{figure}[ht!]
\center
\begin{minipage}[ht!]{0.80\textwidth}
\mbox{\epsfig{file=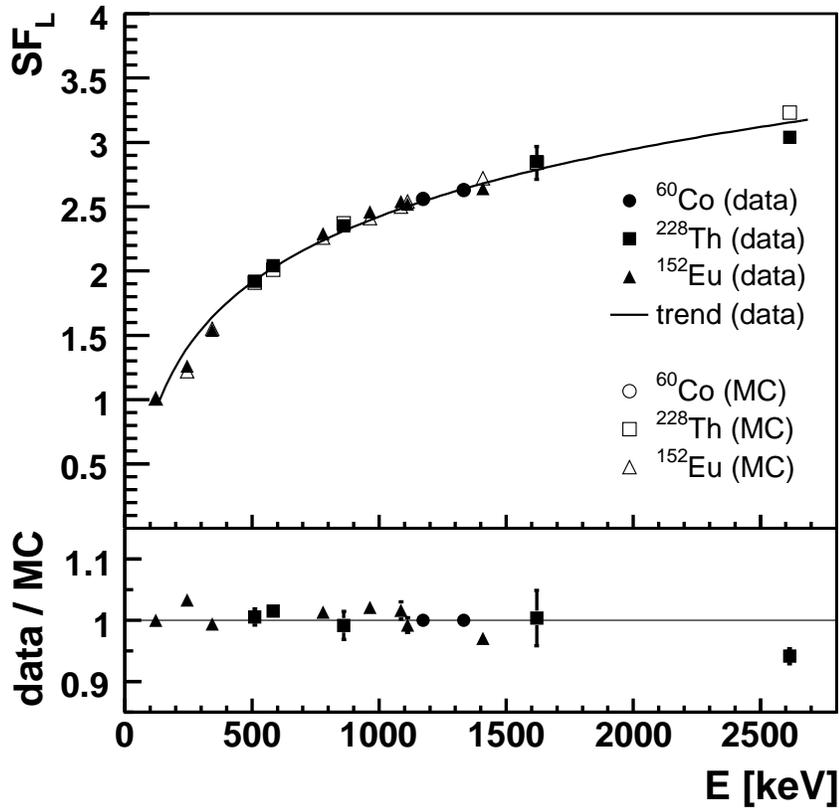,width=\textwidth}}
\end{minipage}
\caption{Top: Line suppression factor as a function of the core energy 
for data (solid marker) and Monte Carlo data (open marker). The trend
curve guides the eye. Bottom: Data to Monte Carlo ratio. The average
deviation is less than 5\%. For a discussion on the agreement between
data and Monte Carlo see section~\ref{subsection:datatomc}.
\label{fig:suppression}}
\end{figure}

The line suppression factors increases from $1.01\pm 0.002$ at 122~keV
to $3.04\pm 0.01$ at $2\,615$~keV, where the suppression increases
with increasing energy. This is expected as the average number of
Compton-scattering processes increases.  Figure~\ref{fig:suppression}
shows the line suppression factors as a function of the core energy
for data and Monte Carlo data. \\

\begin{figure}[ht!]
\center 
\begin{tabular}{cc}
\begin{minipage}[ht!]{0.40\textwidth}
\mbox{\epsfig{file=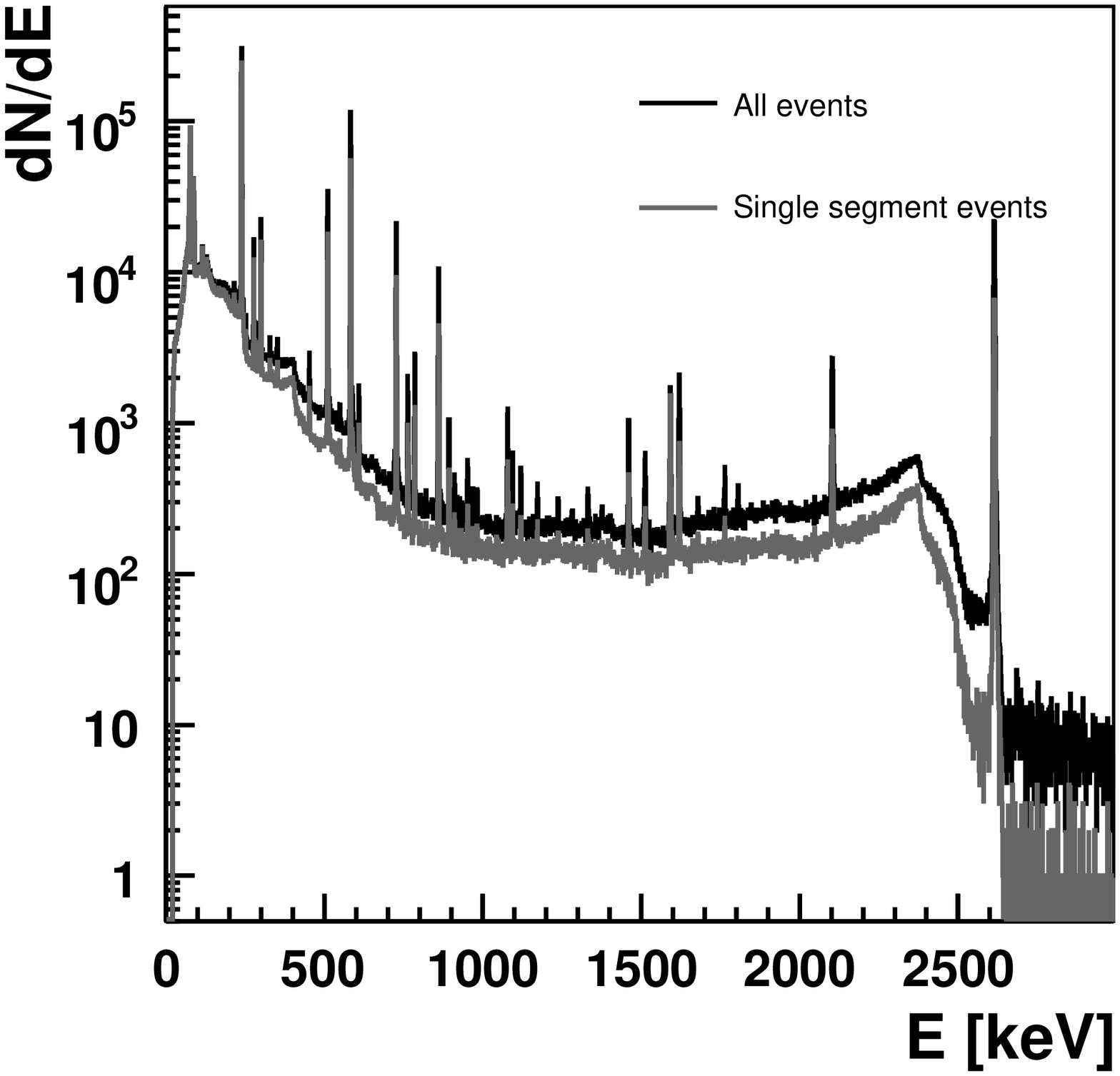,width=\textwidth}}
\end{minipage}
&
\begin{minipage}[ht!]{0.40\textwidth}
\mbox{\epsfig{file=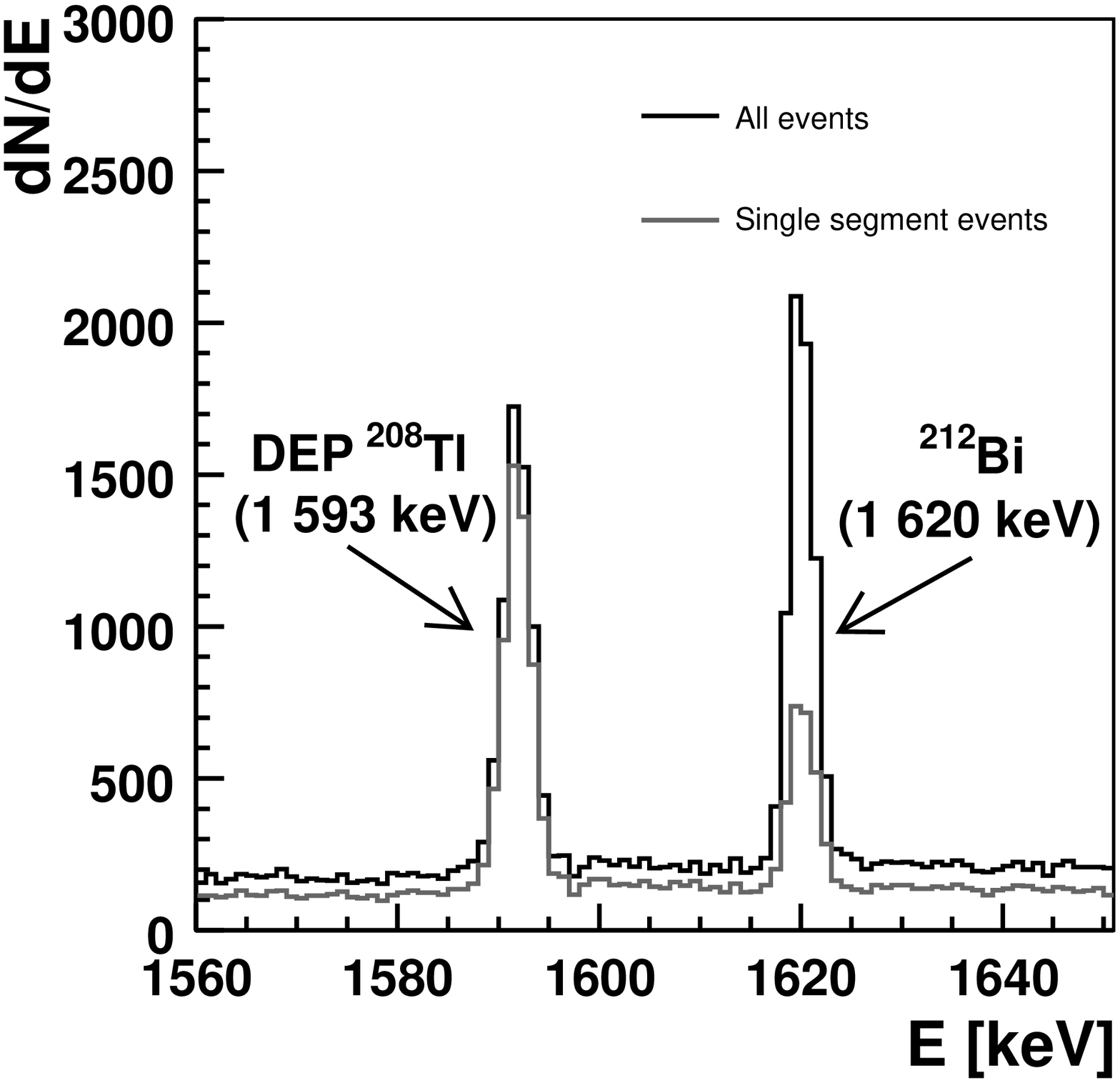,width=\textwidth}}
\end{minipage} \\
\\
\end{tabular}
\caption{Core energy spectrum for the $^{208}$Th source data set for all events 
(black spectrum) and those with a segment multiplicity of
$N_{\mathrm{S}}=1$ (grey spectrum). The left spectrum shows the energy
region up to 3~MeV, the right spectrum is a close-up of the region
around 1.6~MeV. The double escape peak from $^{208}$Tl ($1\,593$~keV)
is hardly suppressed ($SF_{L}=1.09\pm 0.02$) while the $^{212}$Bi line
($1\,620$~keV) is suppressed by a factor of $SF_{L}=2.85\pm
0.01$. Note that background has not been subtracted from the spectra.
\label{fig:suppression_th228}}
\end{figure}

The double escape peak of the $2\,615$~keV photon from the
de-excitation of $^{208}$Tl at $1\,593$~keV is basically not
suppressed. These Class~L events have a very localized energy
deposition. In comparison, the $1\,620$~keV line from the $^{212}$Bi
is suppressed by a factor of $2.85\pm 0.01$. These events are
predominantly encompassed in Class~M.
Figure~\ref{fig:suppression_th228} shows the energy spectrum for the
$^{228}$Th source with and without a segment multiplicity requirement
of $N_{\mathrm{S}}=1$. The left figure shows the energy region up to
3~MeV, the right figure shows a close-up of the region around
1.6~MeV. Note that background has not been subtracted from the data
spectra. \\

The number suppression factor for the $^{60}$Co source is
$SF_{N}=14.2\pm 2.1$. It is large compared to the suppression factor
for the $^{208}$Th source of $SF_{N}=1.68\pm 0.002$. For the latter,
the $Q_{\beta\beta}$ region lies within the Compton continuum of the
$^{208}$Tl photon. A single scattering process can cause a local
energy deposit. In contrast, for the $^{60}$Co source an energy
deposit in this energy region is only possible if both photons
($1\,173$~keV and $1\,333$~keV) deposit energy in the same segment. In
contrast, the number suppression factor for $0\nu\beta\beta$-decay
events is expected to be close to unity as the electrons in the final
state mostly deposit energy in only one segment. \\

\begin{figure}[ht!]
\center 
\begin{tabular}{cc}
\begin{minipage}[ht!]{0.40\textwidth}
\mbox{\epsfig{file=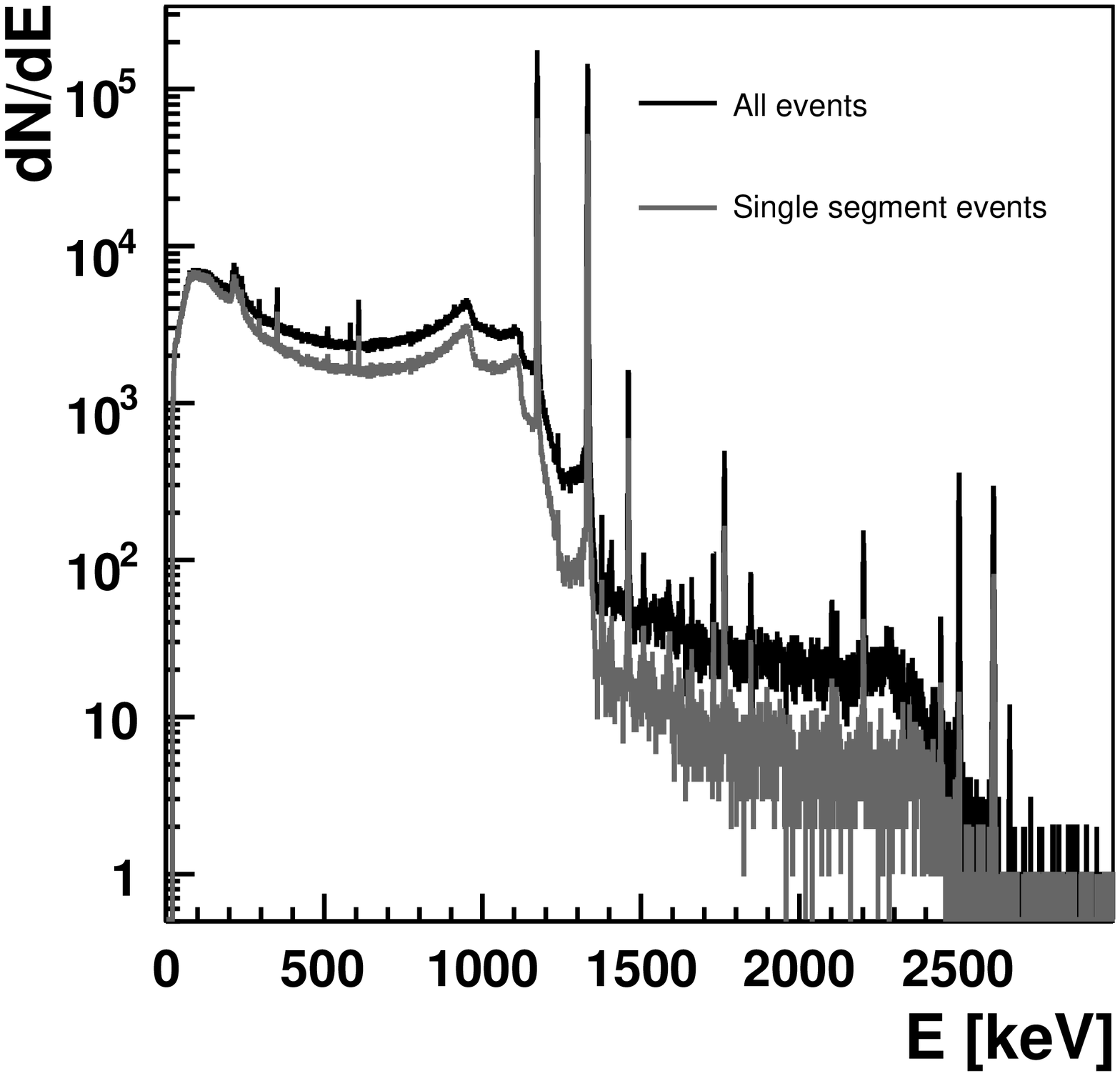,width=\textwidth}}
\end{minipage}
&
\begin{minipage}[ht!]{0.40\textwidth}
\mbox{\epsfig{file=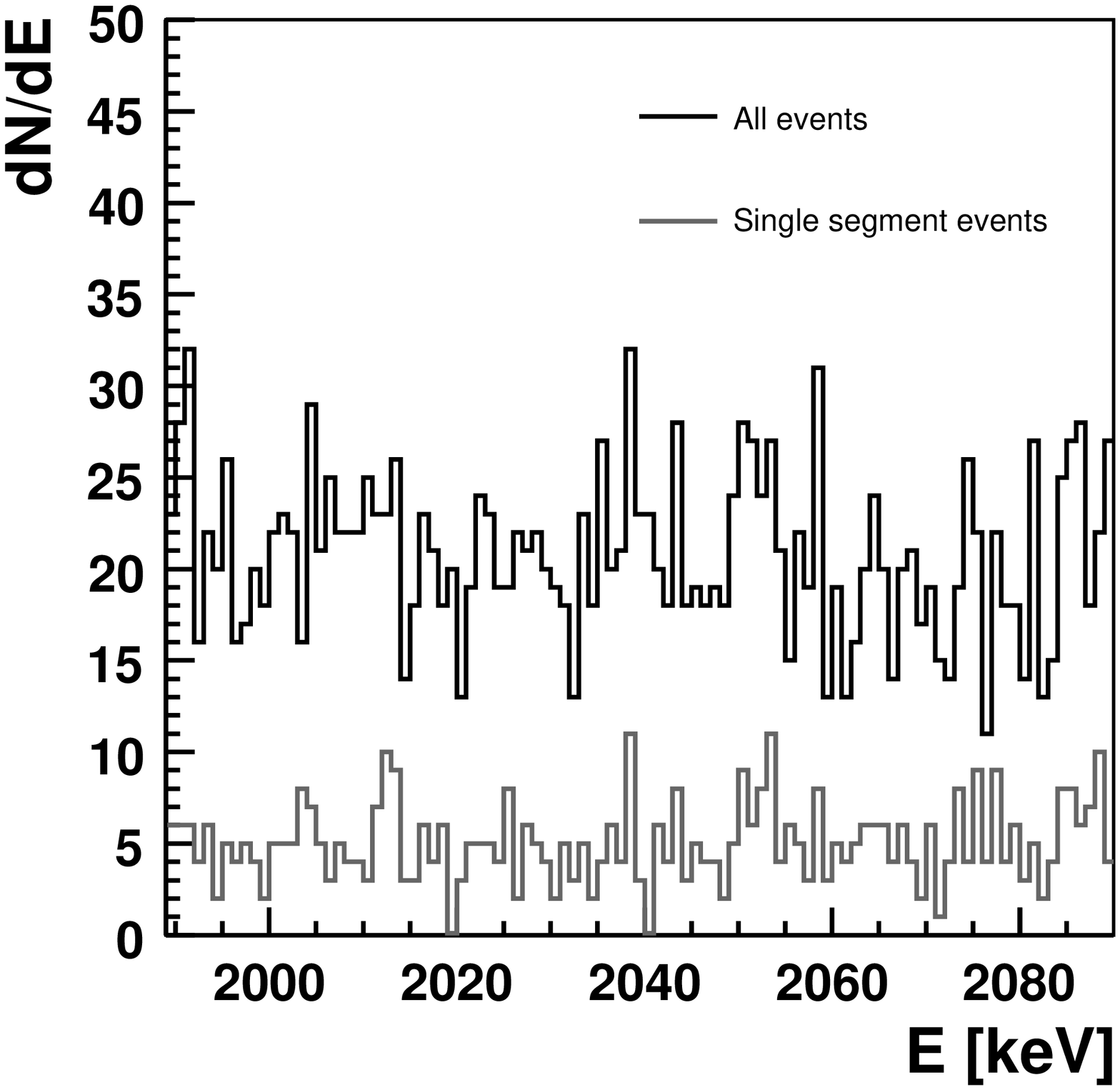,width=\textwidth}}
\end{minipage} \\
\\
\end{tabular}
\caption{Core energy spectrum for the $^{60}$Co source data set for all 
events (black spectrum) and those with a segment multiplicity of
$N_{\mathrm{S}}=1$ (grey spectrum). The left spectrum shows the energy
region up to 3~MeV, the right spectrum is a close-up of the
$Q_{\beta\beta}$-region ($2\,039$~keV). In this energy region, the
single-segment spectrum is suppressed by a factor of $SF_{N}=14.2\pm
2.1$. Note that background has not been subtracted from the spectra.
\label{fig:suppression_co60}}
\end{figure}

Figure~\ref{fig:suppression_co60} shows the energy spectrum for the
$^{60}$Co source with and without a segment multiplicity requirement
of $N_{\mathrm{S}}=1$. The left figure shows the energy region up to
3~MeV, the right figure shows a close-up of the region around the
$Q_{\beta\beta}$-value of $^{76}$Ge. Note that background has not been
subtracted from the spectra. \\

\clearpage 

\subsection{Segmentation study} 
 
In order to study the identification of photons for different
segmentation schemes the segment energies are added in three patterns:
sector, ring and hemisphere. A sector is obtained by adding the
energies of equal $\phi$ segments. A ring is obtained by adding the
energies of equal $z$ segments. Two hemispheres are obtained by adding
the energies of sectors~0,~1,~2 and sectors~3,~4,~5, respectively. The
effective number of segments for the segmentation schemes are 6, 3 and
2, respectively. \\

The suppression factors for each segmentation scheme are obtained as
described in section~\ref{subsection:photonidentification} with
segments replaced by sectors, rings or
hemispheres. Table~\ref{table:suppression_segmentation} shows the
number suppression factor and line suppression factors for selected
photon lines for all four schemes (including the 18-fold segmentation
scheme) obtained from the measurements. As expected, the number and
line suppression factors increase with an increasing effective number
of segments.

\begin{table}[ht!]
\caption{Number and line suppression factors for selected photon lines 
for all four segmentation schemes. An estimated background fraction
has been subtracted from the data. The numbers in brackets are the
effective number of segments for the specific scheme. The
uncertainties are statistical uncertainties only.
\label{table:suppression_segmentation}}
\center
\begin{tabular}{lccccc}
\hline 
Source     & Energy           & $SF$ (18)                            & $SF$ (6)                   & $SF$ (3)                   & $SF$ (2) \\  
           & [keV]            &                                      &                            &                            & \\ 
\hline			      
$^{152}$Eu & \phantom{0\,}344 & \phantom{0}1.54 $\pm$ 0.01           & 1.36 $\pm$ 0.004           & 1.24 $\pm$ 0.003           & 1.12 $\pm$ 0.003  \\ 
$^{60}$Co  & $1\,333$         & \phantom{0}2.63 $\pm$ 0.01           & 1.94 $\pm$ 0.01\phantom{0} & 1.71 $\pm$ 0.004           & 1.30 $\pm$ 0.003 \\ 
$^{228}$Th & $2\,615$         & \phantom{0}3.04 $\pm$ 0.02           & 2.16 $\pm$ 0.01\phantom{0} & 1.86 $\pm$ 0.01\phantom{0} & 1.38 $\pm$ 0.01\phantom{0} \\ 
\hline 			      
$^{60}$Co  & $2\,039$         & 14.2\phantom{0} $\pm$ 2.1\phantom{0} & 9.63 $\pm$ 1.21\phantom{0} & 3.92 $\pm$ 0.33\phantom{0} & 2.61 $\pm$ 0.19\phantom{0} \\ 
$^{228}$Th & $2\,039$         & \phantom{0}1.68 $\pm$ 0.02           & 1.43 $\pm$ 0.02\phantom{0} & 1.40 $\pm$ 0.02\phantom{0} & 1.18 $\pm$ 0.02\phantom{0} \\ 
\hline 
\end{tabular} 
\end{table} 
  
\begin{figure}[ht!]
\center
\begin{minipage}[ht!]{0.80\textwidth}
\mbox{\epsfig{file=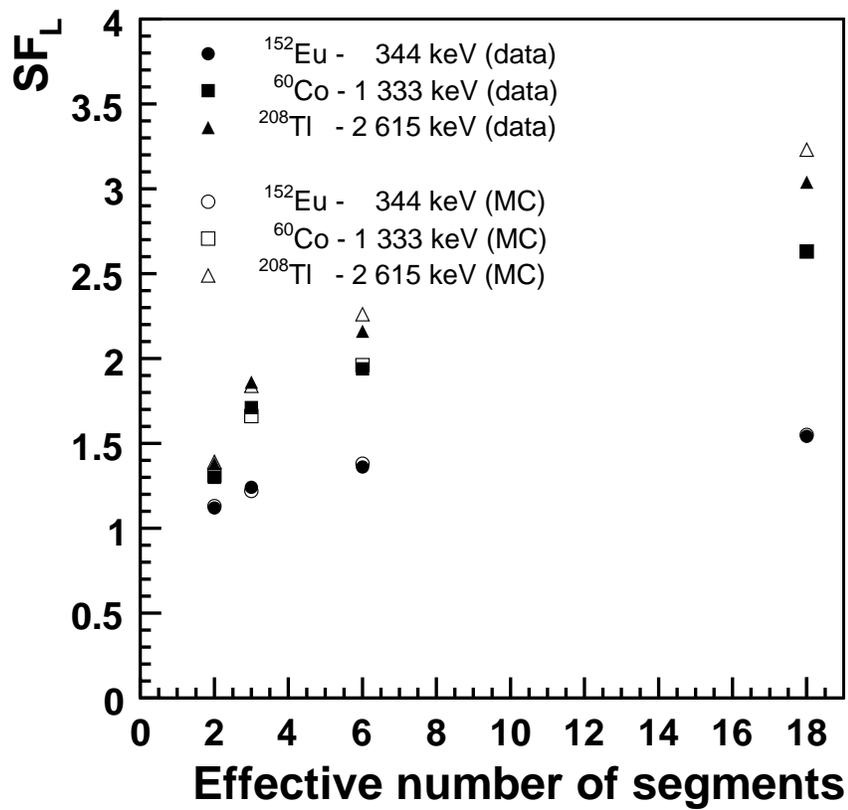,width=\textwidth}}
\end{minipage}
\caption{Line suppression factors for selected photon lines as a function 
of the effective number of segments for data (solid marker) and Monte
Carlo data (open marker).
\label{fig:suppression_segmentation}}
\end{figure}

Figure~\ref{fig:suppression_segmentation} shows the line suppression
factors for the selected photon lines as a function of the effective
number of segments for data and Monte Carlo.

\clearpage 

\subsection{Further studies} 

The effect of the energy threshold on the line suppression factors was
studied by varying the threshold of the core and segment channels from
15~keV to 100~keV. A sharp rise of the line suppression factors is
observed towards 15~keV due to an increased noise level. This behavior
is not observed in Monte Carlo. Between 20~keV and 100~keV the line
suppression factors decrease by up to 7\%. In particular, the double
escape peak at $1\,593$~keV decreases by 4\%. \\
 
A variation of the position of the source was performed. The distance
between the source and the crystal was varied. In addition, the
sources were placed at half the crystal height. The radial distance
between the source and the crystal was varied. No significant
difference in the suppression factors was found. \\
 
The uncertainty on the background fraction in the data samples is
estimated to be 0.1\%. A variation of the estimated background
fraction by this amount did not reveal significant differences in the
line suppression factors. For $^{60}$Co the number suppression factor
decreases with an increasing background fraction because background
events in the energy region around 2~MeV are mostly singly
Compton-scattered photon events from $^{208}$Tl.

\subsection{Data to Monte Carlo comparison} 
\label{subsection:datatomc}  

In the following, the results of the Monte Carlo simulation introduced
in section~\ref{section:simulation} are compared with data for the
$^{60}$Co measurement. Equivalent results are obtained for all three
sources used. \\

\begin{figure}[ht!]
\center 
\begin{tabular}{cc}
\begin{minipage}[ht!]{0.40\textwidth}
\mbox{\epsfig{file=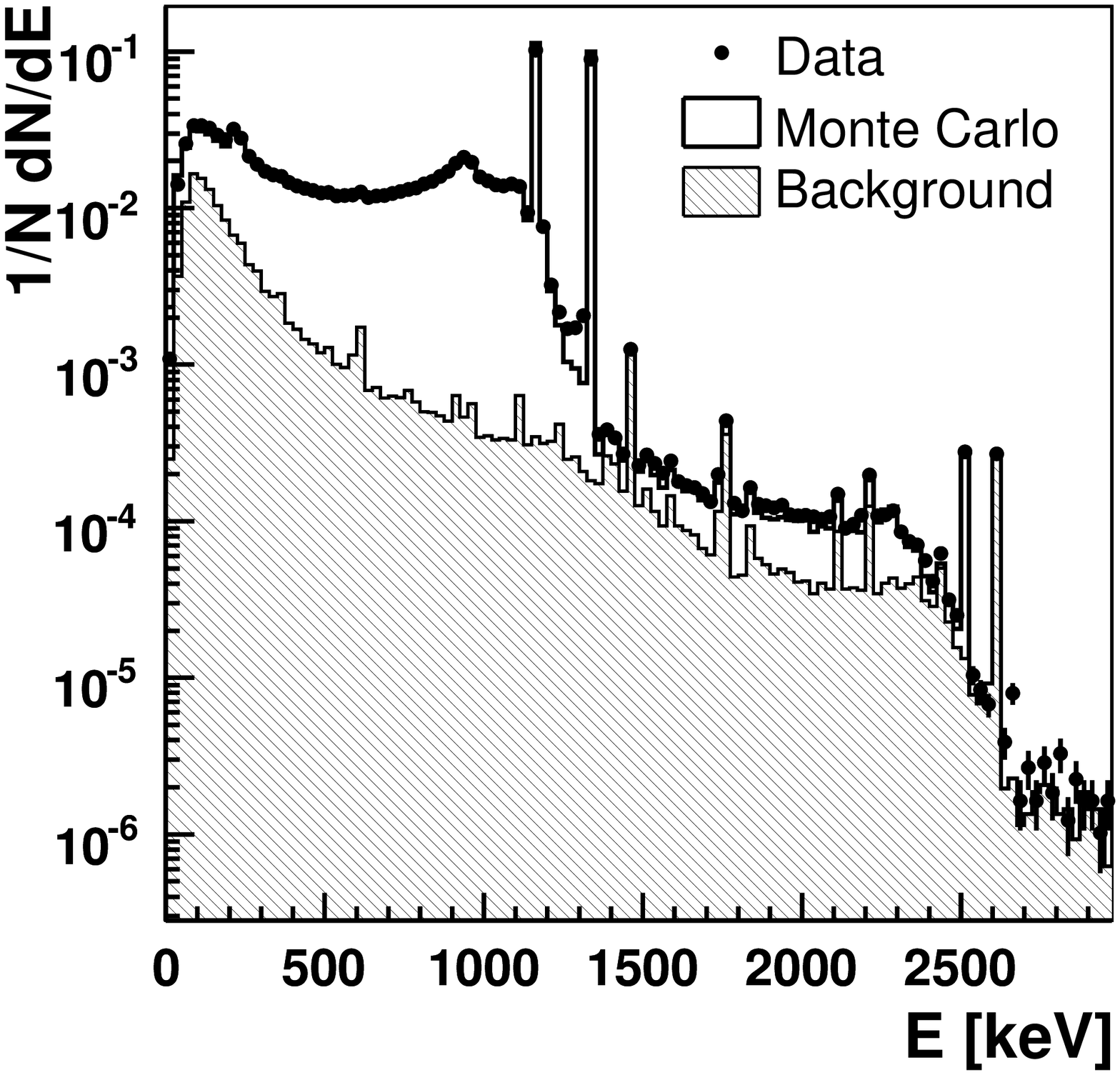,width=\textwidth}}
\end{minipage}
&
\begin{minipage}[ht!]{0.40\textwidth}
\mbox{\epsfig{file=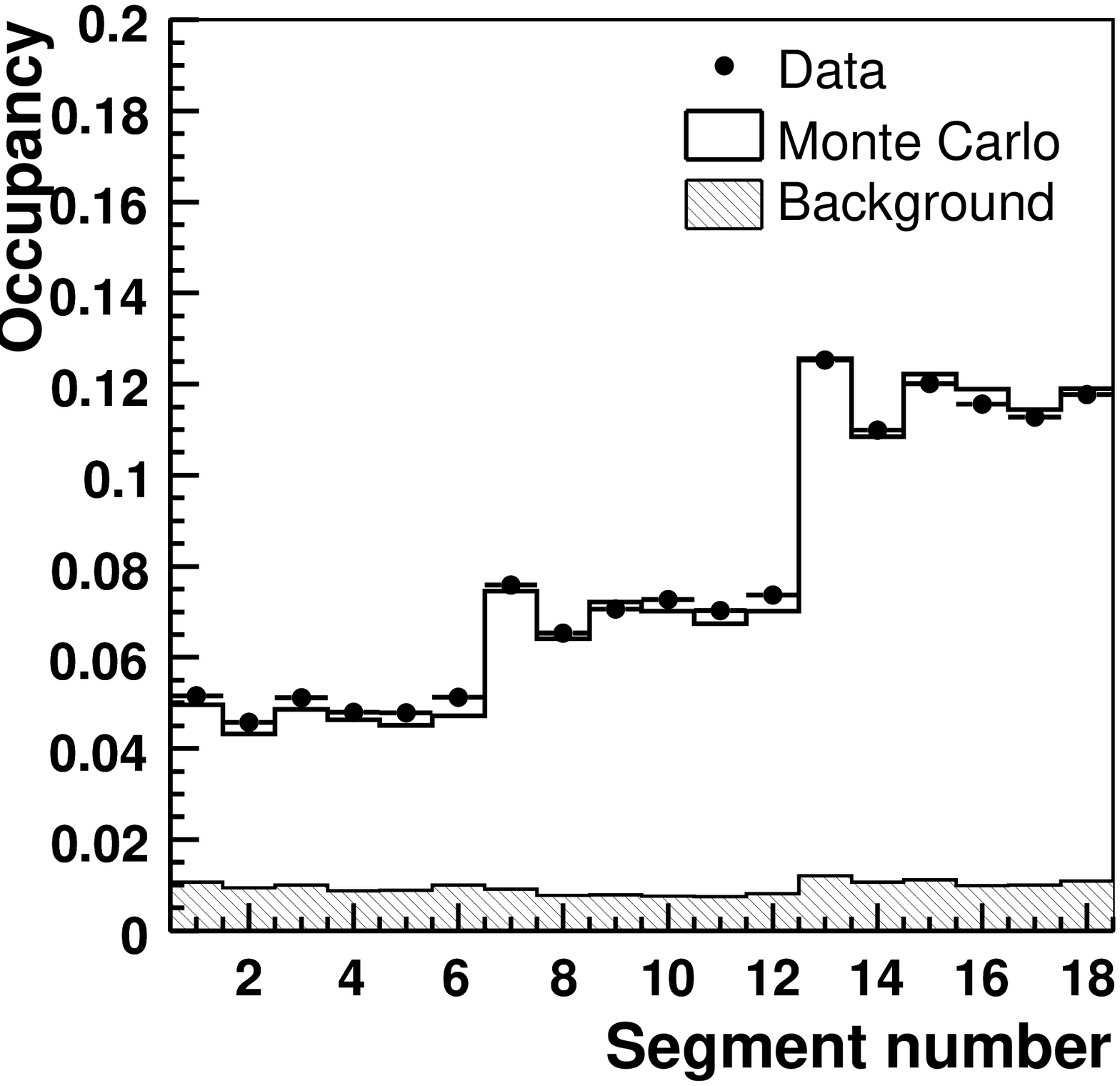,width=\textwidth}}
\end{minipage} \\
\\
\begin{minipage}[ht!]{0.40\textwidth}
\mbox{\epsfig{file=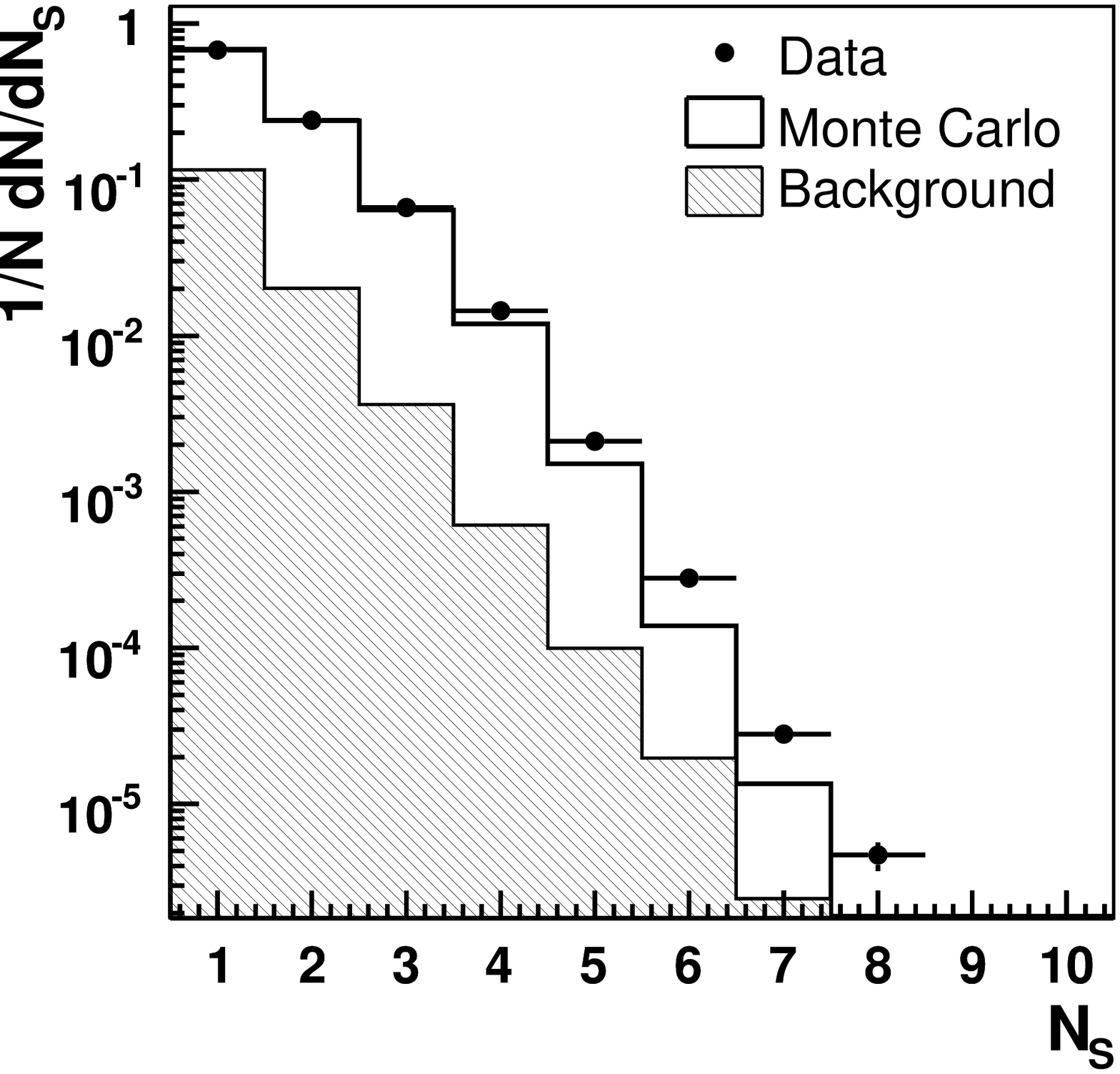,width=\textwidth}}
\end{minipage}
&
\begin{minipage}[ht!]{0.40\textwidth}
\mbox{\epsfig{file=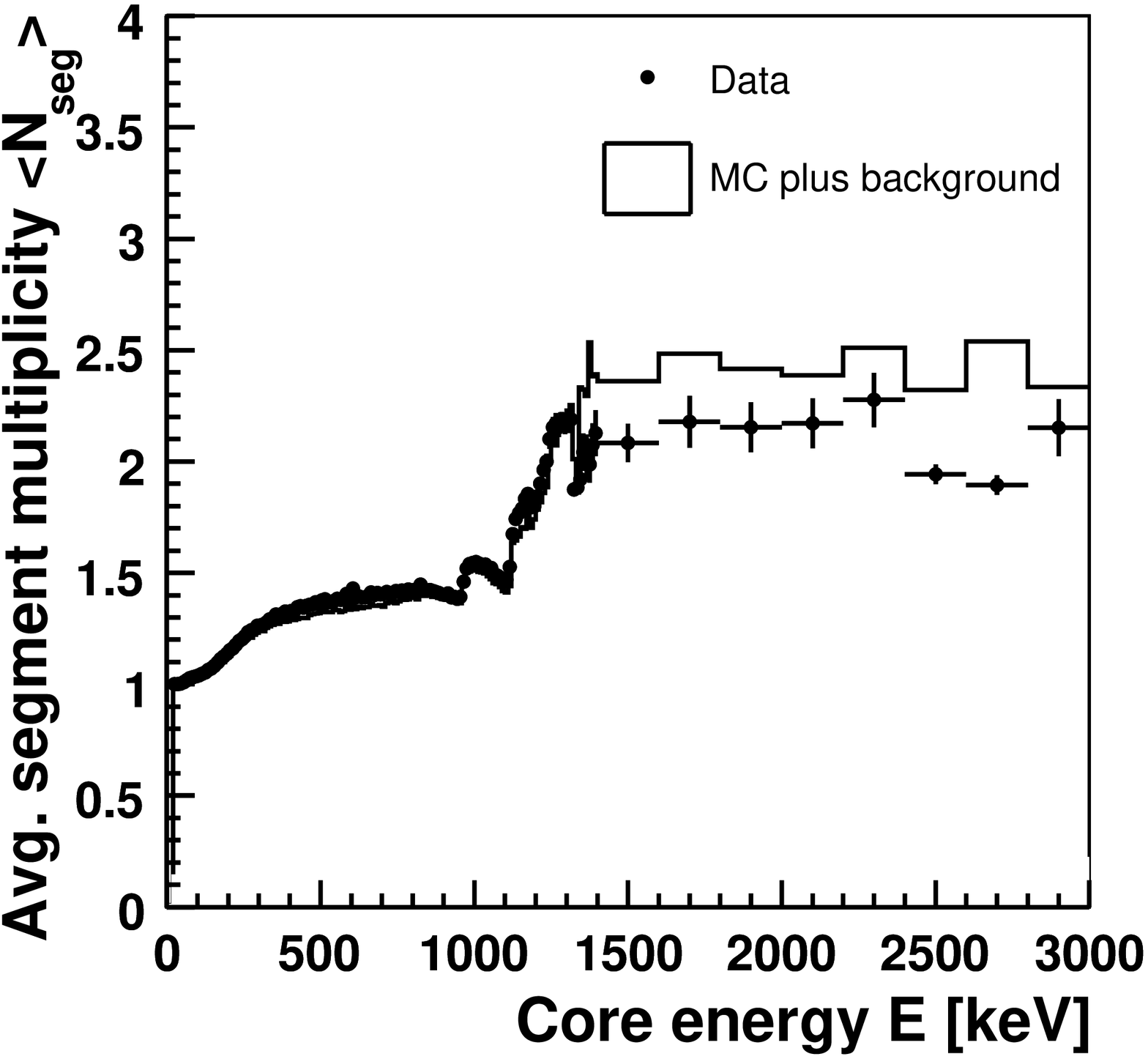,width=\textwidth}}
\end{minipage} \\
\\ 
\begin{minipage}[ht!]{0.40\textwidth}
\mbox{\epsfig{file=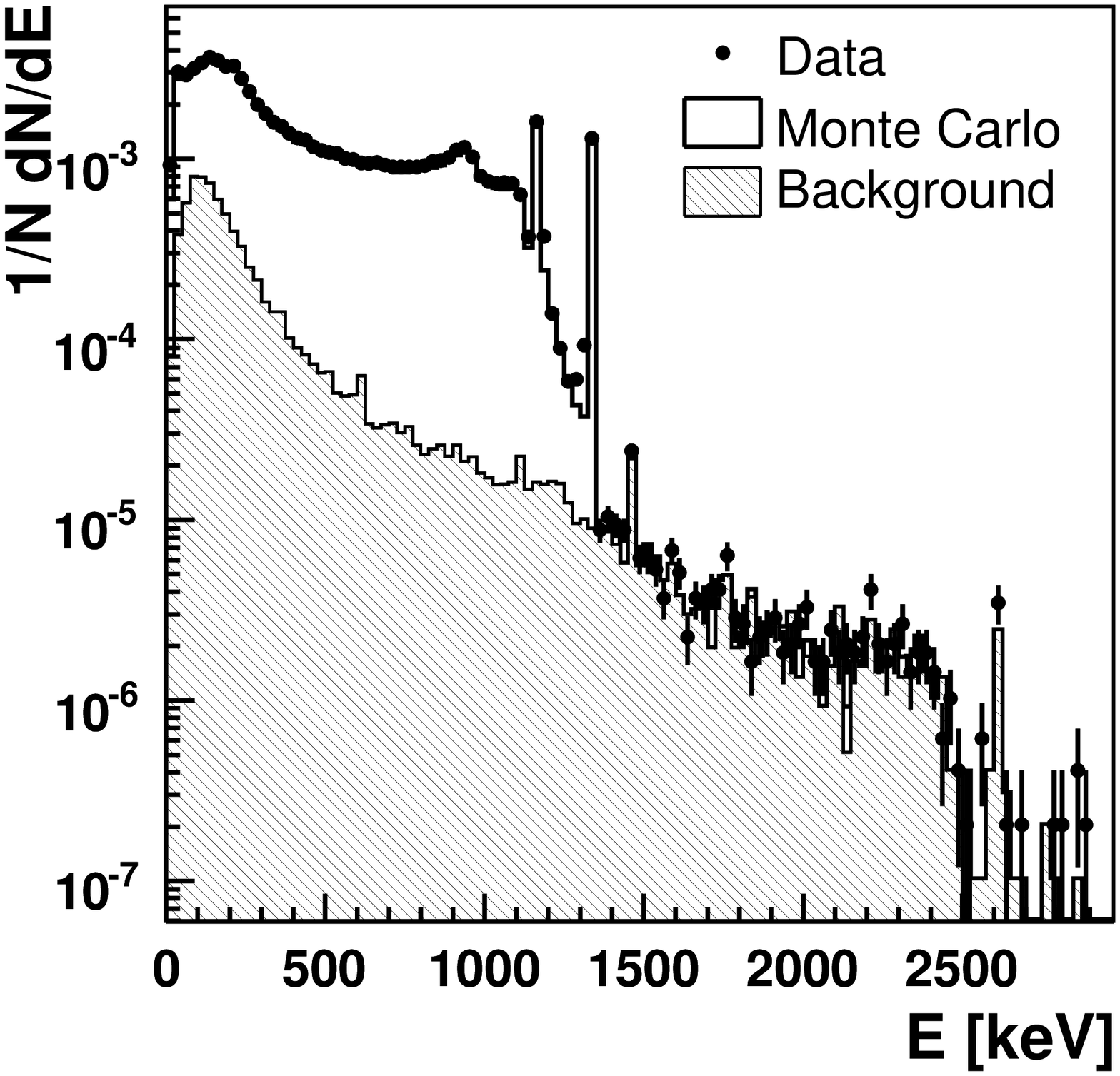,width=\textwidth}}
\end{minipage}
&
\begin{minipage}[ht!]{0.40\textwidth}
\mbox{\epsfig{file=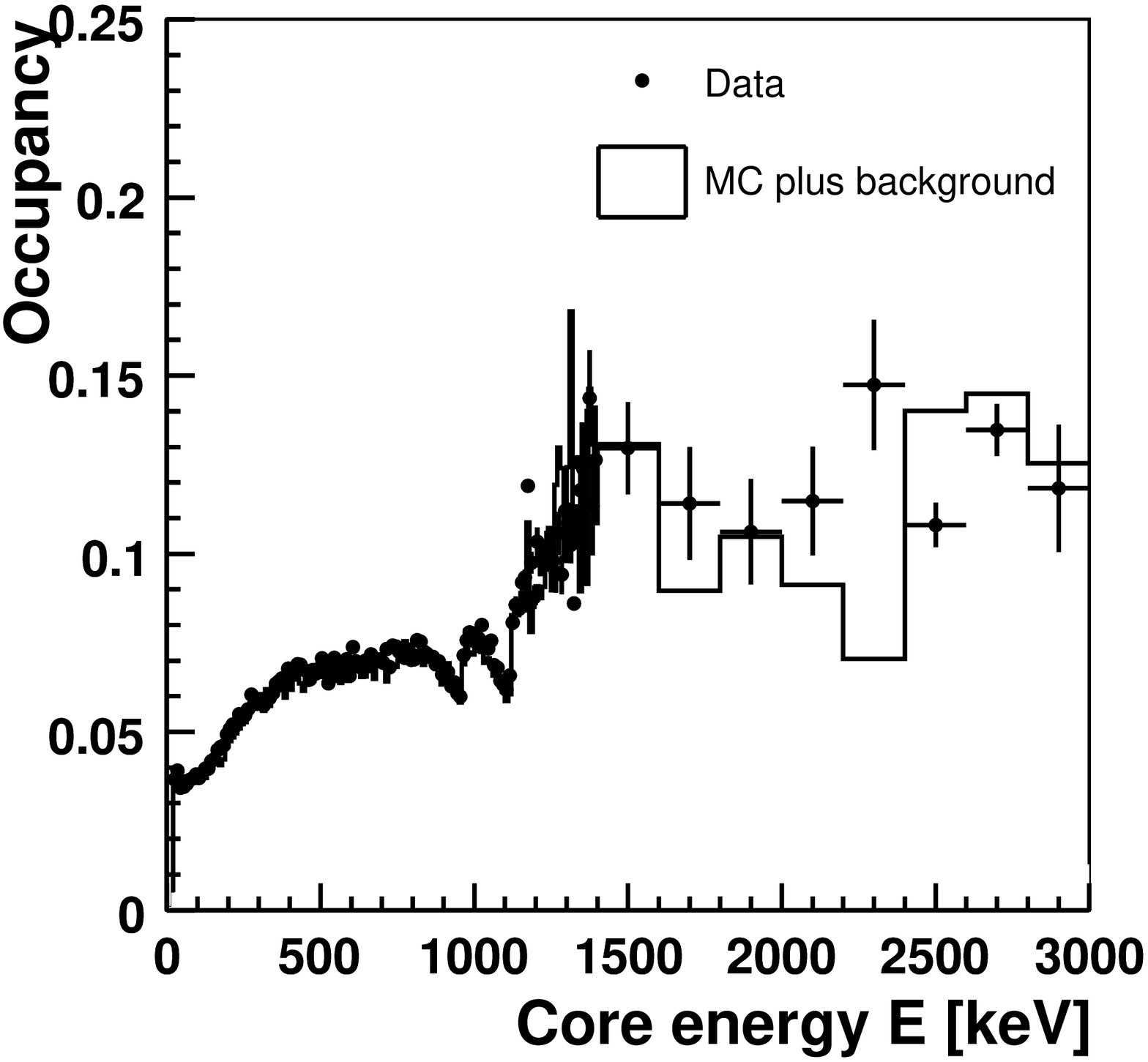,width=\textwidth}}
\end{minipage} \\
\end{tabular}
\caption{Comparison between data and Monte Carlo plus background data for 
several quantities under study for the $^{60}$Co source data set. The
data is indicated by the black marker. The background data is
represented by the hatched histogram, the Monte Carlo data by the open
histogram. The background contribution is estimated as previously
described. Top, left: core energy spectrum. Top, right: occupancy of
all segments. Middle, left: segment multiplicity. Middle, right:
average multiplicity as a function of the core energy. Bottom, left:
energy spectrum taken with segment~10. Bottom, right: occupancy of
segment~10 as a function of core energy. See text for further details.
\label{fig:datatomc}}
\end{figure}

Figure~\ref{fig:datatomc} (top, left) shows the core energy spectrum
for the $^{60}$Co source. The data is indicated by the black
marker. Also shown is the statistical uncertainty. The background data
is represented by the hatched histogram, the Monte Carlo data by the
open histogram. The background contribution is estimated as previously
described. For energies below 100~keV the Monte Carlo plus background
data exceeds the data due to the trigger turn-on which is not
described by Monte Carlo. The Compton continuum of the two $^{60}$Co
lines is described by Monte Carlo with an average deviation of
about~5\%. The number of events under the peak for the two $^{60}$Co
lines are lower in data by 10\%. Tails left and right of the gamma
peaks in data are due to pile-up and not described by the
simulation. The region above 1.3~MeV is dominated by background
events. In this region the average deviation between data and Monte
Carlo plus background data is of the order of 10\% or less. In
particular, the number of events under the peak for the $^{60}$Co
summation line and the $^{208}$Tl line agree within the statistical
uncertainties. \\

Figure~\ref{fig:datatomc} (top, right) shows the occupancy of each
segment, i.e. the fraction of events in which energy is deposited in
the segment under study. No cut on the energy has been
applied. Clearly visible are three groups of segments (1-6, 7-12,
13-18) which correspond to the three z-positions bottom, middle and
top, respectively. As expected, the bottom segments have the lowest,
the top segments the highest occupancy. A pattern within each group is
present which can be explained by the drift anisotropy of the charge
carriers. The structure is reproduced by Monte Carlo using an
effective model for the anisotropy. Without taking the anisotropy into
account no structure is visible. The deviation between data and Monte
Carlo plus background data ranges up to 5-10\%. \\

Figure~\ref{fig:datatomc} (middle, left) shows the segment
multiplicity $N_{\mathrm{S}}$. Data and Monte Carlo range up to
multiplicities of 7-8 with an average multiplicity of 1.4. For
multiplicities up to 3 the deviation between data and Monte Carlo plus
background data ranges up to 5\%. For higher multiplicities the data
exceeds the Monte Carlo with increasing multiplicity. \\
 
Figure~\ref{fig:datatomc} (middle, right) shows the average segment
multiplicity as a function of the energy measured with the core
electrode up to 3~MeV. For energies up to 1~MeV the average
multiplicity increases with energy from 1 to about 1.5. For energies
between 1~MeV and 1.3~MeV the multiplicity increases up to 2.2. The
deviation between data and Monte Carlo plus background data for
energies below 1.3~MeV ranges up to 10\%. For higher energies the
average deviation is of the order of 15\%, where the Monte Carlo plus
background data shows a larger average multiplicity. \\
 
Figure~\ref{fig:datatomc} (bottom, left) shows the energy spectrum
measured with (arbitrarily chosen) segment~10 up to energies of
3~MeV. The features described for the core electrode are also seen
here. The deviation between data and Monte Carlo plus background data
ranges up to 10\%. \\
 
Figure~\ref{fig:datatomc} (bottom, right) shows the occupancy for
segment~10 as a function of the energy measured with the core
electrode up to 3~MeV. The occupancy ranges from 5\% to 8\% for
energies below 1.3~MeV. For larger energies the occupancy ranges up to
15\%. The deviation between data and Monte Carlo plus background data
ranges up to 10\%. \\

The suppression factors derived from the data are compared with those
yielded from the Monte Carlo simulation in
Table~\ref{table:suppression}. The average deviation between data and
Monte Carlo data is less than 5\%. The absolute values depend on the
DAQ-inefficiency. \\

The overall agreement between data and Monte Carlo plus background
data is good. The remaining discrepancy between data and Monte Carlo
plus background data could stem from (1) the modeling of the exact
detector geometry including dead layers and segment borders, (2) the
missing modeling of the drift of charge carriers, especially close to
the surface of the crystal, (3) the missing simulation of the
pre-amplifier response, (4) effects which are not accounted for in the
simulation such as pile-up and trigger turn-on in the DAQ system and
(5) the missing angular correlation between the photons of the
cascades\footnote{Preliminary studies show that the angular
correlation between the photons emitted in the decay of $^{60}$Co has
no significant effect on the suppression factors.}. 

\clearpage 

%

\section{Conclusions and outlook}
\label{section:conclusions}

A study with a segmented {\sc GERDA} prototype detector was
performed. It was shown that the identification of events with
multiply scattered photons in the final state using segmented
detectors is feasible. The reduction of events in which only the full
photon energy is deposited inside the detector is energy dependent and
increases from $1.01\pm0.002$ at~122~keV to $3.13\pm0.01$ at
$2\,615$~keV in the experimental setup described. The suppression of
Compton-scattered events in the $Q_{\beta\beta}$-region of $^{76}$Ge
($2\,039$~keV) coming from $^{60}$Co and $^{228}$Th sources was
measured to be $14.2\pm2.1$ and $1.68\pm0.02$, respectively. \\
 
An additional study showed that the identification of photon events is
improved by increasing the number of segments and that it is stable
with respect to energy threshold variations and the background
estimate. \\
 
In comparison to the suppression factors calculated
in~\cite{segmentation} the suppression for a single crystal is lower
than for an array of detectors due to the geometrical acceptance.
 
A data to Monte Carlo comparison, considering background from
radioactive isotopes in the laboratory, showed an agreement with
deviations on the level of 5-10\%. This shows that simulations based
on the {\sc MaGe} tool are suitable and can reliably predict
background reductions as presented in~\cite{segmentation}. \\
 
The identification of photonic events can be further improved by the
analysis of the time structure of the detector responses. This is
currently under study.


\section{Acknowledgements}
 
The authors would like to thank the {\sc GERDA} and {\sc Majorana}
Monte~Carlo groups for their fruitful collaboration and cooperation on
the {\sc MaGe} project.


\addcontentsline{toc}{section}{Bibliography}
%


\end{document}